\documentclass[prl,twocolumn,showpacs,aps,floatfix,amsmath,amssymb,amsfonts]{revtex4-1}

\usepackage{amsmath}
\usepackage{graphicx}
\usepackage{color}
\usepackage{pdfpages}

\newcommand{\beq}{\begin{equation}}
\newcommand{\eeq}{\end{equation}}

\newcommand{\rs}{\rm \scriptscriptstyle}

\begin{document}
\title{Three-body interaction of Rydberg slow light  polaritons}
\author{Krzysztof Jachymski}
\author{ Przemys\l aw Bienias}
\author{ Hans Peter B\"{u}chler}
\affiliation{
Institute for Theoretical Physics III  and Center for Integrated Quantum Science and Technology, University of Stuttgart, Pfaffenwaldring 57, 70550 Stuttgart, Germany}
\date{\today}

\begin{abstract}
We study a system of three photons in an atomic medium coupled to Rydberg states near the conditions of electromagnetically induced transparency. Based on the analytical analysis of the microscopic set of equations in the far-detuned regime,  the effective three-body interaction for these Rydberg polaritons is derived. For slow light polaritons, we find a strong three-body repulsion with the remarkable property that three polaritons can become essentially non-interacting at short distances. This analysis allows us to derive the influence of the three-body repulsion on bound states and correlation functions of photons propagating through a one-dimensional atomic cloud.\end{abstract}

\maketitle

Quantum systems consisting of a few interacting bodies are a central point of attention in different fields of physics~\cite{Glockle2012,Zinner2014}. Despite their apparent simplicity, in general few-body problems are not analytically solvable and posses fascinating emergent properties. A prominent example is the existence of universal three-body bound states for bosons with pairwise short-range interactions discovered by Efimov~\cite{Efimov1970}. In addition, three-body forces~ can have strong influence on the properties of quantum many-body systems such as nuclear systems~\cite{Brown1969}, neutron stars~\cite{Steiner2012}, and fractional quantum Hall states 
~\cite{Greiter1991}.  It is thus natural to look for systems in which three-body interactions could be controlled for the purpose of quantum simulations. Several proposals have been made in this context, mainly utilizing ultracold atoms and molecules~\cite{Buchler2007,Johnson2009,Daley2009,Mazza2010,Petrov2014}. In this Letter, we demonstrate that strong three-body interactions naturally appear between Rydberg slow light polaritons.

Rydberg slow light polaritons have recently emerged as a promising approach to engineer a strong interaction between photons \cite{Friedler2005,Gorshkov2011,Peyronel2012,Parigi2012,Firstenberg2013}. It is based on the combination of 
 electromagnetically induced transparency (EIT)~\cite{Fleischhauer2005} and the strong interaction between Rydberg states.
Under EIT conditions, single photons propagate in the medium as dark polaritons with reduced velocity and significant admixture of the Rydberg state~\cite{Fleischhauer2000}. 
Then, the strong interactions between Rydberg atoms that give rise to the blockade effect~\cite{Lukin2001,Heidemann2007} can be mapped onto polaritons, resulting in effective interaction potential~\cite{Gorshkov2011,Bienias2014,Maghrebi2015}. The sign, strength and range of the interactions can be tuned by varying the Rabi frequencies and detuning of the lasers as well as principal quantum number of the atoms. The Rydberg EIT scheme has been used to study quantum nonlinear optics at single photon level~\cite{Peyronel2012,Dudin2012,Maxwell2013,Hoffman2013,Firstenberg2013,Gorniaczyk2014,Tiarks2014} and can be applied to realize strongly correlated many-body states of light~\cite{Otterbach2013,Gorshkov2013,Maghrebi2015a,Moos2015,Weber2015,Sommer2015,Ningyuan2015}.  However, the analysis of these systems has so far been restricted to models based on the effective two-body interaction between the polaritons. 


In this Letter, we make an essential step towards studying strongly interacting many-body systems made of slow light polaritons. Basing on the microscopic set of equations describing photons in an EIT medium in the far-detuned regime, we analytically derive the interaction potential for a three-body system and demonstrate the appearance of a strong three-body interaction potential in addition to the previously discussed effective two-body potential. We find that especially in the experimentally interesting regime of slow light polaritons, the influence of the three-body interaction can be equally important as the contribution from the effective two-body interaction.  This strongly influences the properties of three-body bound states as well as the correlation function of three photons propagating through a realistic one-dimensional setup, allowing for simulation of exotic many-body models and creation of strongly correlated multiphoton states.

\begin{figure*}[t]
\centering
\includegraphics[width=0.95\textwidth]{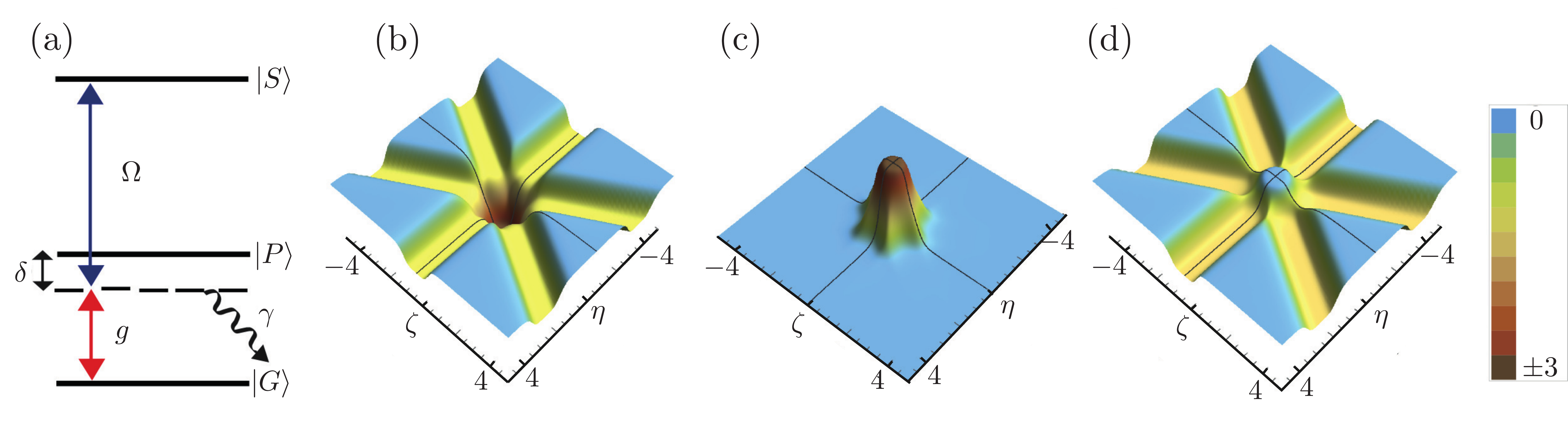}
\caption{\label{fig1} (a) Level diagram for the Rydberg EIT setup. (b) The two-body contribution $\sum_{i<j} V_{\rs eff}^{(2)}(x_i-x_j)$ to  the total interaction potential $U$  in Jacobi coordinates. Energies are expressed in units of $1/|\chi|$, while  lenghts are expressed in units of the blockade radius $\xi$. For illustration, we choose a real valued $\Delta$  with $C_6 \Delta<0$, and set $\alpha=1$. (c) The pure three-body interaction $
V_{\rs eff}^{(3)}$. (d) Total interaction potential  $ U$ with three-body and two-body contributions, which demonstrates that in this regime three polaritons become non-interacting at short distances.}
\end{figure*}

We start with the microscopic derivation of the three-body interaction potential between the slow light polaritons. The atomic medium consists
of three-level atoms with $|G\rangle$ being the ground state, and an intermediate state $\left|P\right>$ coupled to a Rydberg level $\left|S\right>$ by a control laser with Rabi frequency $\Omega$ and detuning $\Delta$; the latter includes the decay of the intermediate level by $\Delta = \delta -i \gamma$, see Fig.~\ref{fig1}. The probe photons are tuned near the EIT condition and therefore photons entering the atomic medium are converted into slow light polaritons with a large admixture of the Rydberg state. 
The effective two-polariton interaction potential has been derived by several different approaches before \cite{Gorshkov2011,Firstenberg2013,Bienias2014,Sommer2015}. The conceptually simplest approach is based on the analysis for a single photonic mode realized in a single mode cavity: the stationary Schr\"{o}dinger equation reduces for two photons  in the cavity to a set of coupled equations for different components of the wave function. Solving these equations~\cite{Sommer2015,SupMat} determines the energy shift  in the presence of two photons in the cavity, and relates directly to the two-polariton interaction potential.

For large $|\Delta| \gg \Omega$, which will be assumed throughout this manuscript, the intermediate level can be adiabatically eliminated. The effective  interaction potential takes the form~\cite{Gorshkov2011,Firstenberg2013,Bienias2014,Sommer2015}
\beq
V_{\rs eff}^{(2)}({\bf r})=\alpha^2 \frac{V({\bf r})}{1-\chi \: V({\bf r})},
\eeq
with  $\chi=\Delta/(2\hbar \Omega^2)$ and the van der Waals interaction $V({\bf r})=C_6/|{\bf r}|^6$  between the Rydberg states. Furthermore,  $\alpha=g^2/(g^2+\Omega^2)$ denotes the probability to find a single polariton in the Rydberg level with $g$ being the collective atom-photon coupling. We note that the interactions are saturated at short distances as a result of the Rydberg blockade mechanism. The characteristic length scale for this process, called the blockade radius, is defined as $\xi=|C_6 \chi|^{1/6}$.

One can expect that for more than two photons higher order terms in $\alpha$ can arise, which then correspond to effective many-body interactions. Here, we are interested in the three-body term. The derivation is again most conveniently performed in a single mode cavity with three photons present in the system, and expressing the system in terms of the stationary Schr\"{o}dinger equation for the photons and the atomic matter. The analysis is presented in detail in the supplementary material~\cite{SupMat}. The important step in the derivation is the assumption that the size of the photonic mode is much larger than the blockade radius, which is equivalent to the condition of low energies. Then, the effective interaction can be read off from the analytical result for the small energy shift for the photons in the cavity and decomposed into a sum of two-body and three-body contributions. 
The pure three-body interaction term  takes the form
\beq
V_{\rs eff}^{(3)}({\bf x}_1, {\bf x}_2, {\bf x}_3)\!=\!\alpha^{3}\sum_{i<j} \frac{V_3({\bf x}_1, {\bf x}_2, {\bf x}_3)\!-\! V({\bf x}_i\!-\!{\bf x}_j)}{1- \chi V({\bf x}_i\!-\!{\bf x}_j)}  ,
\label{eq:3bp}
\eeq
with
\begin{equation}
V_3({\bf x}_1, {\bf x}_2, {\bf x}_3)=\frac{\sum_{i<j}{V({\bf x}_i-{\bf x}_j)}}{3-2  \chi \sum_{i<j}{V({\bf x}_i-{\bf x}_j})}.
\end{equation}
It immediately follows that the three-body interaction exhibits opposed behavior at short distances with respect to the two-body one: while $V_{\rm eff}^{(2)}({\bf r})$ saturates at $ -\alpha^2/\chi$, the three-body interaction exhibits the opposite sign and saturates at $+ 3 \alpha^3/\chi$. As expected, the three-body interaction is suppressed in $\alpha$ for weak coupling of the photons $\alpha \ll 1$, but exhibits an equal strength as the effective two body interaction for slow light polaritons with $g \gg \Omega$.

From now on, we will measure lengths in units of the blockade radius $\xi$ and energies in units of  $1/|\chi|=2\hbar\Omega^2/|\Delta|$. While our derivation is general and can be applied regardless of the geometry of the system, the influence of the three-body interaction is most conveniently studied for a one-dimensional setup on which we will now focus. 
By introducing the Jacobi coordinates defined as $R=(x_1+x_2+x_3)/\sqrt{3} \xi$,  $\eta=(x_1-x_2)/ \sqrt{2} \xi$, and $\zeta=\sqrt{2/3}((x_1+x_2)/2-x_3)/\xi$, the center of mass $R$ disappears from Eq.~\eqref{eq:3bp}, and therefore the three-body interaction depends only on the two relative coordinates $\eta$ and $\zeta$ as  shown in Fig.~\ref{fig1}. We note that the six-fold symmetry which is naturally present for three particles interacting via two-body forces is preserved by our three-body term. 
It is remarkable that the saturation of the full interaction potential at short distances takes the from  $-3 \alpha^3 \Omega^2/g^2 \chi$, and vanishes in the limit of slow light 
$g\gg \Omega$ with $\alpha =1$.  Then, dissipative losses from the decay of the intermediate $|P\rangle$~level, which are accounted for by complex value of $\Delta$, are suppressed.  
A simple explanation of this behavior can be obtained by the following argument: at short distances, the Rydberg blockade enables only a single Rydberg excitation. Then, the value of the effective potential at short distances for $n$ polaritons is determined by the probability to find one Rydberg excitation and $(n-1)$ photons, i.e., $n g^2 \Omega^{2 (n-1)}/(g^2+\Omega^2)^n$ multiplied by the dispersive energy  shift for the  $n-1$ photons due to their coupling to the $|P\rangle$~level. The latter reduces to $-(n-1)g^2 \hbar/\Delta $.
%
%
This simple estimation provides indeed the correct saturation for two  and three polaritons.


%
%
%
%
%
%
%
%

We can now extend the analysis to the full propagation problem of polaritons in a one-dimensional setup as studied 
experimentally in Ref.~\cite{Firstenberg2013}. For simplicity, we focus on the coherent dynamics and neglect the losses 
from spontaneous emission, i.e., $\gamma=0$, in the regime $C_6 \Delta <0$. The effective low energy Hamiltonian for the polaritons reduces to $H= H_{\rs kin} + U$ with the interaction $U$ including the two-body interaction as 
well as the three-body interaction
\begin{equation}
U(x_1,x_2,x_3)  = \sum_{i < j}  V_{\rs eff}^{(2)}(x_{i}-x_{j}) + V_{\rs eff}^{(3)}(x_{1},x_{2},x_{3}).
\end{equation}
In turn, the kinetic energy  $H_{\rs kin}$ of the polaritons is well accounted for by the expansion of the dispersion relation at low momenta
providing the slow light velocity $v_{g} = \Omega^2/(g^2+\Omega^2) c$ and a mass term \cite{Gorshkov2011,Otterbach2013,Bienias2014} 
\begin{equation}
H_{\rs kin} = \sum_{j=1}^{3}\left[ i \hbar v_{g} \partial_{x_{j}} + \frac{\hbar^2}{2 m} \partial_{x_j}^2\right]  
\end{equation}
with $m =\hbar (g^2+\Omega^2)^3/ (2 c^2 g^2 \Delta \Omega^2)$. It is important to stress that the only approximation in deriving the Hamiltonian $H$ is the restriction to the low energy regime, i.e., $|E_{j}|<  \hbar \Omega^2 /|\Delta|$ with $E_{j}$ the energy of the polaritons.




\begin{figure}
\centering
\includegraphics[width=0.48\linewidth]{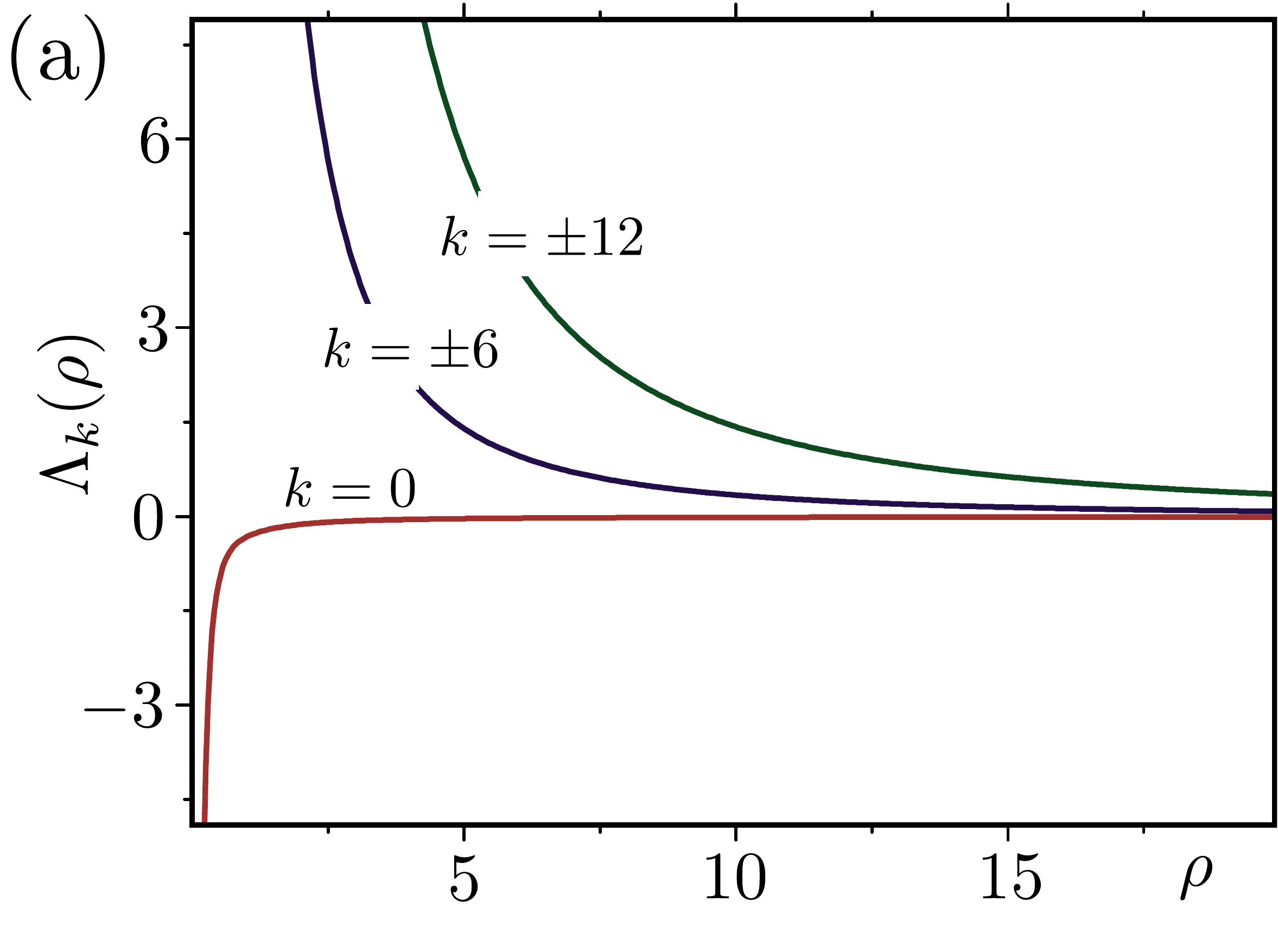}
\includegraphics[width=0.49\linewidth]{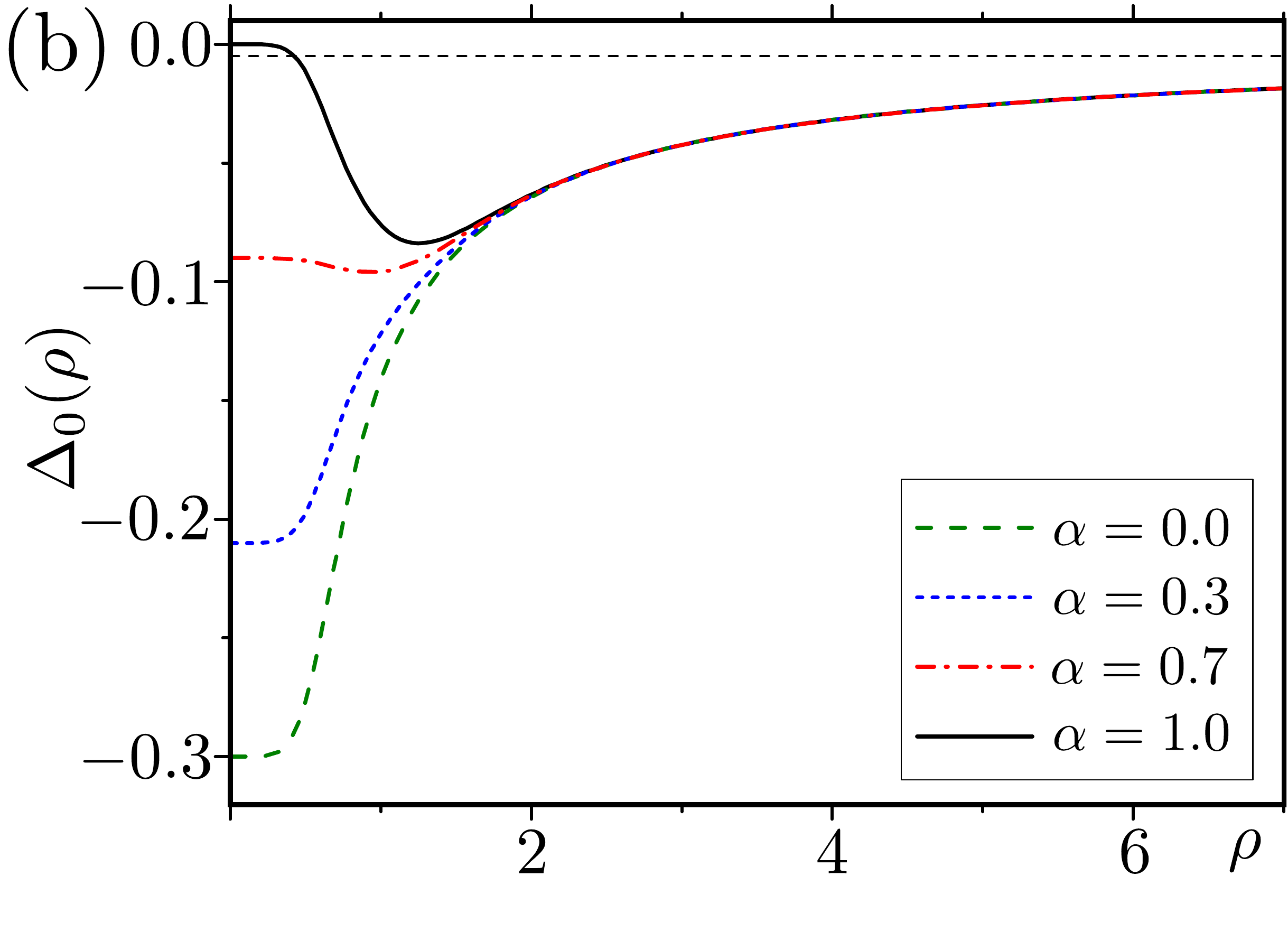}
\caption{\label{fig:ad}(a) Adiabatic curves $\Lambda_k(\rho)$ for $\alpha=1$ and $\lambda=0.1$. Note, that only partial waves with a difference in angular momentum by a multiple of 6 are coupled to each other. (b) Lowest effective adiabatic potential $\Delta_0(\rho)$ for different values of $\alpha$ and $\lambda=0.1$. The horizontal thin dashed line denotes the energy of the two-body bound state.}
\end{figure}

We will now analyze how the short-range repulsion affects the properties of the system, and first focus on the three-body bound state. 
In Jacobi coordinates, the center of mass motion can be separated, and the Hamiltonian describing the relative motion of the polaritons reduces to a two-dimensional problem and can be conveniently written as
\beq
\label{eq:ham}
H_{\rm rel}=-\frac{\partial^2}{\partial\eta^2}-\frac{\partial^2}{\partial\zeta^2}+\lambda  \tilde{U}(\eta,\zeta),
\eeq
with $\tilde{U} = \chi U/\alpha^2$ and $\lambda=|\alpha^2 m\xi^2/(\hbar^2\chi)|$. Note, that in the far detuned regime, the two-polariton potential is always attractive and its strength is
determined by the dimensionless parameter $\lambda$, which can also be written as $\lambda = \kappa_{\xi}^2  (\Omega^2+g^2)/ g^2  $ with $\kappa_{\xi} = \xi g^2/|\Delta| c$ the off-resonant optical thickness per blockade radius. Note, that for weak interactions with $\lambda \ll 1$, the two-body potential is well described by an attractive $\delta$-function potential, while for increasing interacitons
with $\lambda > 1$ the bound state energies become comparable to $\hbar \Omega^2/\Delta$, and we start to leave the low energy regime.
An exact solution for the bound states of a three-body system with pairwise $\delta$-function interactions shows a single three-body bound state with energy $-4B$, where $B$ is the binding energy of the two-body bound state~\cite{Mcguire1964}. In our case, the repulsive three-body interaction will increase the energy of this three-body bound state. 

%
%

In order to study the properties of the full system, we first make use of the adiabatic potentials method, which has proven successful for pairwise delta interactions~\cite{Gibson1987,Mehta2005}. To this end, we introduce the hyperspherical coordinates $\rho$, $\theta$ with $\eta=\rho\sin\theta$, $\zeta=\rho\cos\theta$. We then expand the wave function into partial waves $\psi=\sum_k{\frac{\Phi_k(\rho)}{\sqrt{\rho}}\frac{{\rm exp}(ik\theta)}{\sqrt{2\pi}}}$, 
which provides a set of coupled radial equations. Diagonalization of these equations at fixed position provides the adiabatic potentials; note, that each channel is still dominated by the corresponding partial wave, and therefore we keep the index $k$.
Within the adiabatic approximation, we can neglect the coupling terms between the channels~\cite{SupMat}. This results in a set of one-dimensional equations of the form
\beq
\left(-\frac{d^2}{d\rho^2}+\frac{k^2-1/4}{\rho^2}+\Delta_k(\rho)\right)\Phi_k(\rho)=E\Phi_k(\rho),
\eeq
where $\Delta_k$ is the effective interaction in channel $k$, equivalent to total adiabatic potential curve $\Lambda_k=\Delta_k+(k^2-1/4)/\rho^2$, see Fig.~\ref{fig:ad}.
\begin{figure}
   \centering
\includegraphics[width=0.45\textwidth]{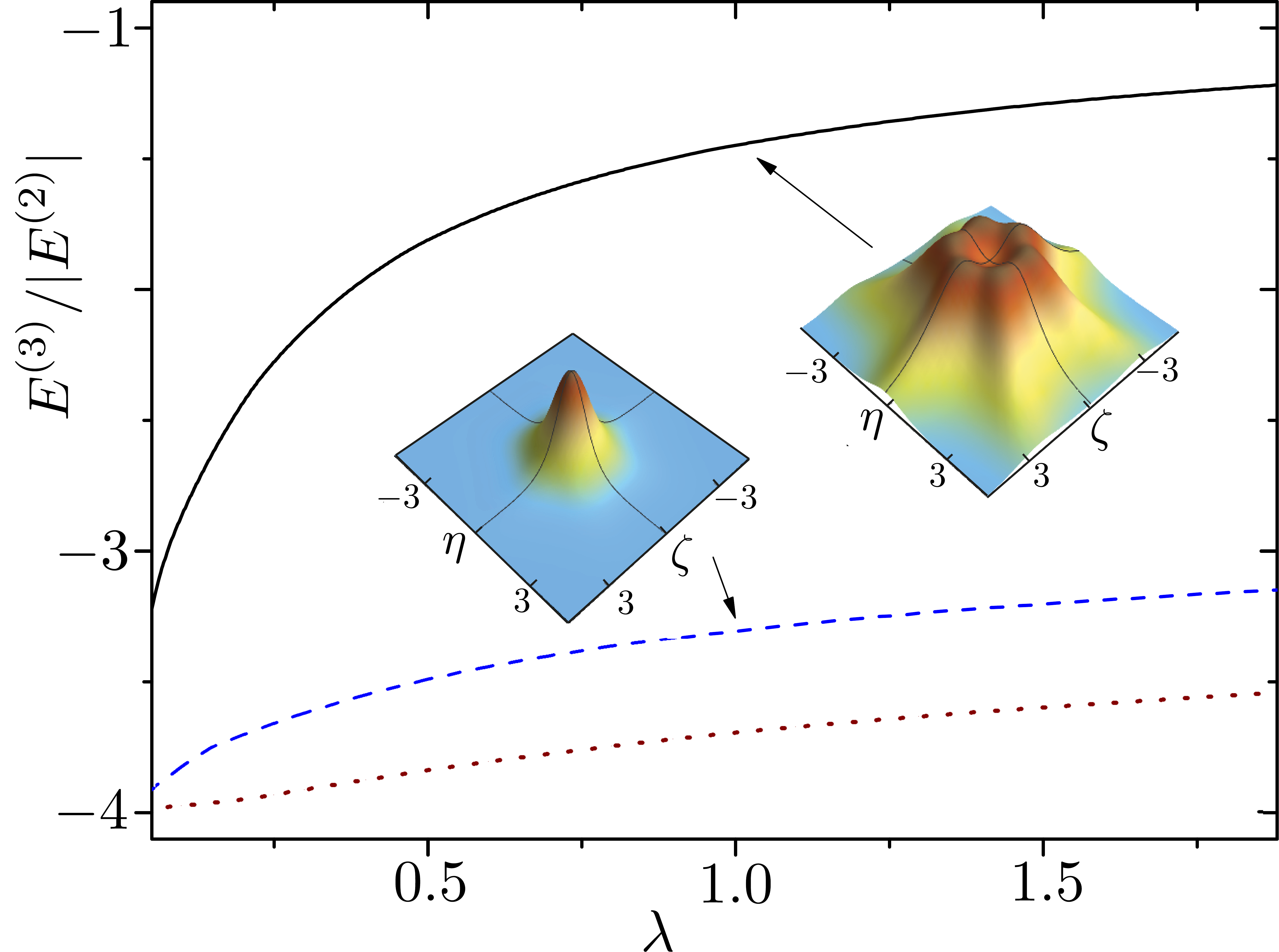}
		\caption{\label{fig:3bb}Binding energy $E^{(3)}$ of the three-body bound state in units of the corresponding two-body enegy $E^{(2)}$ as a function of $\lambda$. The dotted  (red) line corresponds to the situation of pure two-body interaction, while modifications due to the three-body interaction are shown for weak  ($\alpha=0.1$, blue dashed line) and strong ($\alpha=1$, black solid line) three-body repulsion. The insets show the wave function of the bound state for $\lambda=1$ for two cases.}
\end{figure}
%
Only the lowest ($k=0$) channel is attractive and can support bound states. Furthermore, the lowest curve at large $\rho$ approaches the energy of the two-body bound state; the latter behavior is well understood, as the atom-dimer continuum should start exactly at the energy of this bound state. The impact of the three-body forces becomes more clear when the angular term is subtracted: the lowest effective potential is plotted in the right panel of Fig.~{\ref{fig:ad}} for different values of $\alpha$ but fixed interaction strength $\lambda=0.1$. In the absence of three-body repulsion the potential looks similar to the two-body interaction, with characteristic short-range saturation. When the repulsion is turned on, the short distance behavior changes. However, the potential exhibits an attractive well for any $\alpha\in[0,1]$ regardless of the value of $\lambda$, so a three-body bound state is always expected to exist. Its properties, however, may be strongly dependent on $\alpha$. 

Further insight into the problem can be gained by direct numerical diagonalization of Eq.~(\ref{eq:ham})~\cite{SupMat}. We indeed find the three-body bound state to be the ground state of the system for any value of the parameters, in agreement with the adiabatic approach.  In Fig.~\ref{fig:3bb}, we show the dependence of the energy $E^{(3)}$ for  the three body bound state on $\lambda$ for different values of $\alpha$; it is convenient to show the ratio between the three-body bound state energy   $E^{(3)}$ and the corresponding two-body bound state energy $E^{(2)}$. For $\alpha =0$, we recover the analytical result  of~\cite{Mcguire1964} in the limit of weak interactions. 
%
Three-body repulsion not only shifts the bound state energy, but also provides a significant broadening of the wave function as well as the appearance of characteristic dip at the center; see the inset of~Fig.~\ref{fig:3bb}. 

\textit{Experimental implications:}
We now discuss the detection and the impact of three-body interactions in experiments with photons propagating through a 1D medium. In a realistic situation the photons are injected into the medium in a coherent state with low mean number of photons and the detection  takes place after they leave the medium. Time-resolved measurements give access to the intensity correlation functions. Here we are interested in the third order correlations, which should contain information about the three-body bound state. In the following, we choose parameters, which are close to the experimental parameters achieved in Ref.~\cite{Firstenberg2013}: the condition of slow light with $g \gg \Omega$ implies $\alpha \approx 1$, while the experimentally observed interaction strength expressed in our parameters correspond to $\lambda \approx 0.1$. 

\begin{figure}
\includegraphics[width=0.95\linewidth]{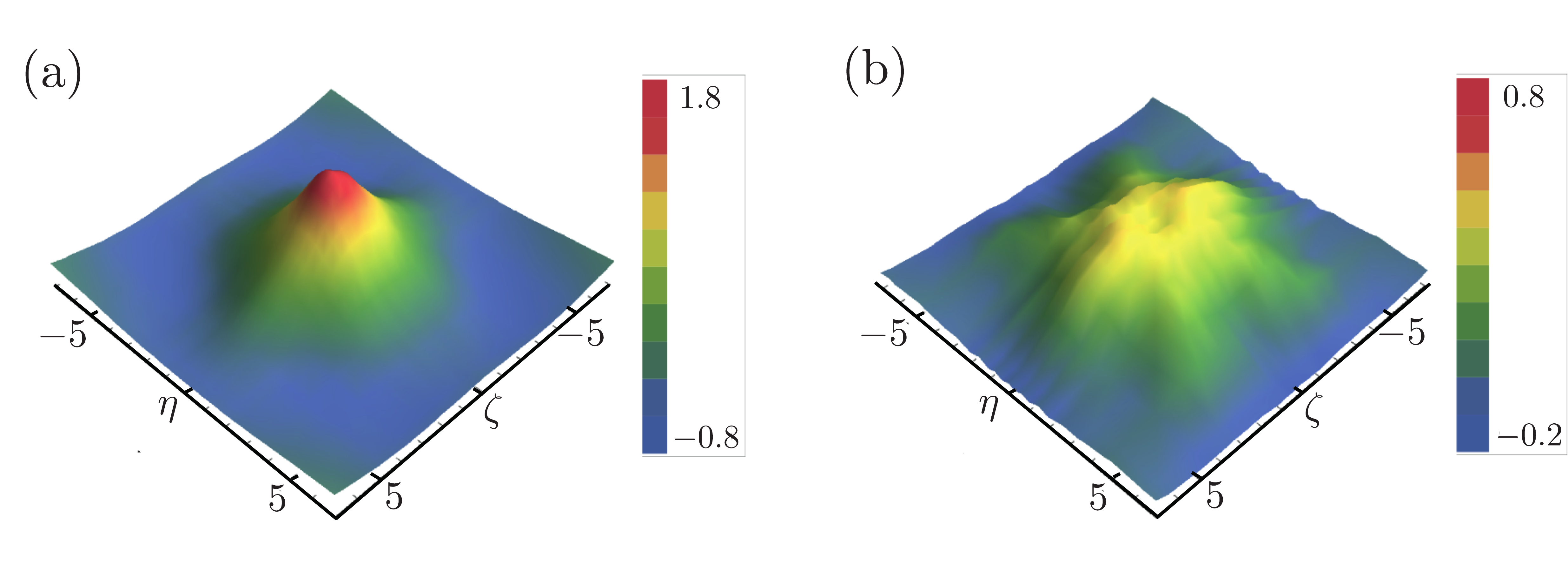} 
\caption{\label{fig:cor}Central peak of the connected part of third order correlation function $\tilde{g}^{(3)}$ for $\lambda =0.1$ and a length of the media $R=20 \xi$. (a) In absence of the three-body interaction,  $\tilde{g}^{(3)}$  shows the characteristic feature of the three-body bound state determined by a $\delta$-function interaction. (b) The full behavior including the strong three-body interaction with $\alpha=1$, demonstrating the characteristic behavior of the three-body bound  state in this regime. Note, that  $\tilde{g}^{(3)}$ approaches zero for large distances.}
\end{figure}

Solving the full propagation problem for three photons is in general extremely challenging.  Therefore, we will here perform a simplified analysis, which has previously turned out to be very successful for two photons \cite{Firstenberg2013}. It is based on the approximation that the atomic medium is a homogeneous slab and that the three-photon component of the wave function obeys the boundary condition $\psi(R=0,\eta,\zeta)=\kappa^3$, where $\kappa$ is the amplitude of the coherent state. Then we have $g^{(3)}(R=0,\eta,\zeta)=1$. This can be decomposed into contributions from bound and scattering states. During the propagation different eigenstates pick up different phases, which leads to formation of a characteristic pattern in the correlation function. For second order correlations, the contribution from the bound state becomes clearly visible~\cite{Firstenberg2013}. To extract information about pure three-body correlations, we note that when one particle is separated from the other two, $g^{(3)}$ approaches the value of $g^{(2)}$. It is thus natural to study the connected part of the correlation function $\tilde{g}^{(3)}$ instead of $g^{(3)}$, which is defined as
\[
\tilde{g}^{(3)}(x_1,x_2,x_3)=2+g^{(3)}(x_1,x_2,x_3)-\!\sum_{i<j}{g^{(2)}(x_i,x_j)}.
\]
The connected correlation function $\tilde{g}^{(3)}(x_1,x_2,x_3)$ obeys the property that it
approaches zero at large particle separation. The numerical determination of $\tilde{g}^{(3)}$
is then straightforward using the full set of eigenstates of Eq.~(\ref{eq:ham}) as a basis set: first, we expand the incoming wave function in this basis; then, each eigenstate acquires a phase during the propagation through the medium proportional to its energy and the distance $R$. This eventually determines
the outgoing wave function and we can compute $\tilde{g}^{(3)}(x_1,x_2,x_3)$ from the final state. 
The result for  $\tilde{g}^{(3)}$  is shown in Fig.~\ref{fig:cor} after propagation distance of $20\xi$ in the medium. The central peak in the correlation functions originates from the bound states and exhibits a stable and characteristic shape; in analogy to the two-body correlation function $g^{(2)}$ \cite{Firstenberg2013}.
We clearly see that for $\alpha=1$ the width of the peak is significantly greater as compared to the absence of three-body interactions, and its shape follows the three-body bound state wave function. This implies that measurement of $\tilde{g}^{(3)}$ should indeed give access to the structure of three-body bound states.

In conclusion,
we have shown that Rydberg polaritons naturally exhibit three-body interactions which strongly affects their few-body properties in the regime of slow light. The short-range three-body repulsion modifies the energy and shape of the three-body bound states of polaritons propagating through a one-dimensional channel, which can be detected in experiments 
by measuring third order correlations. It is a remarkable property that in the regime of slow light with $\alpha \approx 1$, the total interactions vanish at short distances. This creates a region in which three closely lying photons are protected from dissipation.  We therefore expect that the transmission in the dissipative regime for three photons is strongly enhanced.

Our results are independent of the dimensionality of the system and can also be applied to multimode optical cavities. This paves the way to use polaritons for simulating exotic few- and many-body models.
Especially, it is possible to quench the $s$-wave scattering length 
by tuning the parameter $\lambda$ to control the number of bound states in the attractive two-body potential; in analogy to the 1D situation \cite{Bienias2014}. Then, the remaining interaction is dominated by the  repulsive three-body interaction, enabling realization of purely three-body interacting systems of photons in arbitrary dimensions, leading to interesting quantum states of matter; the most prominent example being the Pfaffian states~\cite{Greiter1991}. Other potential applications include creating correlated photonic states by propagating multi-photon pulses through Rydberg EIT medium.

{\it Note}: During the review process, we became aware of a recent related work by Gullans {\it et al}~\cite{Gullans2016}.


We acknowledge fruitful discussions with Maxim Efremov.
This work was supported by the European Union under the ERC consolidator grant SIRPOL (grant N. 681208), the Foundation for Polish Science within the START program, the Alexander von Humboldt Foundation, 
and the Deutsche Forschungsgemeinschaft (DFG) 
within the SFB/TRR 21.

\bibliography{Polrefs}

\begin{thebibliography}{41}%
\makeatletter
\providecommand \@ifxundefined [1]{%
 \@ifx{#1\undefined}
}%
\providecommand \@ifnum [1]{%
 \ifnum #1\expandafter \@firstoftwo
 \else \expandafter \@secondoftwo
 \fi
}%
\providecommand \@ifx [1]{%
 \ifx #1\expandafter \@firstoftwo
 \else \expandafter \@secondoftwo
 \fi
}%
\providecommand \natexlab [1]{#1}%
\providecommand \enquote  [1]{``#1''}%
\providecommand \bibnamefont  [1]{#1}%
\providecommand \bibfnamefont [1]{#1}%
\providecommand \citenamefont [1]{#1}%
\providecommand \href@noop [0]{\@secondoftwo}%
\providecommand \href [0]{\begingroup \@sanitize@url \@href}%
\providecommand \@href[1]{\@@startlink{#1}\@@href}%
\providecommand \@@href[1]{\endgroup#1\@@endlink}%
\providecommand \@sanitize@url [0]{\catcode `\\12\catcode `\$12\catcode
  `\&12\catcode `\#12\catcode `\^12\catcode `\_12\catcode `\%12\relax}%
\providecommand \@@startlink[1]{}%
\providecommand \@@endlink[0]{}%
\providecommand \url  [0]{\begingroup\@sanitize@url \@url }%
\providecommand \@url [1]{\endgroup\@href {#1}{\urlprefix }}%
\providecommand \urlprefix  [0]{URL }%
\providecommand \Eprint [0]{\href }%
\providecommand \doibase [0]{http://dx.doi.org/}%
\providecommand \selectlanguage [0]{\@gobble}%
\providecommand \bibinfo  [0]{\@secondoftwo}%
\providecommand \bibfield  [0]{\@secondoftwo}%
\providecommand \translation [1]{[#1]}%
\providecommand \BibitemOpen [0]{}%
\providecommand \bibitemStop [0]{}%
\providecommand \bibitemNoStop [0]{.\EOS\space}%
\providecommand \EOS [0]{\spacefactor3000\relax}%
\providecommand \BibitemShut  [1]{\csname bibitem#1\endcsname}%
\let\auto@bib@innerbib\@empty
\bibitem [{\citenamefont {Gl{\"o}ckle}(2012)}]{Glockle2012}%
  \BibitemOpen
  \bibfield  {author} {\bibinfo {author} {\bibfnamefont {W.}~\bibnamefont
  {Gl{\"o}ckle}},\ }\href@noop {} {\emph {\bibinfo {title} {The quantum
  mechanical few-body problem}}}\ (\bibinfo  {publisher} {Springer Science \&
  Business Media},\ \bibinfo {year} {2012})\BibitemShut {NoStop}%
\bibitem [{\citenamefont {Zinner}(2014)}]{Zinner2014}%
  \BibitemOpen
  \bibfield  {author} {\bibinfo {author} {\bibfnamefont {N.~T.}\ \bibnamefont
  {Zinner}},\ }\href@noop {} {\bibfield  {journal} {\bibinfo  {journal}
  {Few-Body Systems}\ }\textbf {\bibinfo {volume} {55}},\ \bibinfo {pages}
  {599} (\bibinfo {year} {2014})}\BibitemShut {NoStop}%
\bibitem [{\citenamefont {Efimov}(1970)}]{Efimov1970}%
  \BibitemOpen
  \bibfield  {author} {\bibinfo {author} {\bibfnamefont {V.}~\bibnamefont
  {Efimov}},\ }\href@noop {} {\bibfield  {journal} {\bibinfo  {journal}
  {Physics Letters B}\ }\textbf {\bibinfo {volume} {33}},\ \bibinfo {pages}
  {563} (\bibinfo {year} {1970})}\BibitemShut {NoStop}%
\bibitem [{\citenamefont {Brown}\ and\ \citenamefont
  {Green}(1969)}]{Brown1969}%
  \BibitemOpen
  \bibfield  {author} {\bibinfo {author} {\bibfnamefont {G.}~\bibnamefont
  {Brown}}\ and\ \bibinfo {author} {\bibfnamefont {A.}~\bibnamefont {Green}},\
  }\href@noop {} {\bibfield  {journal} {\bibinfo  {journal} {Nuclear Physics
  A}\ }\textbf {\bibinfo {volume} {137}},\ \bibinfo {pages} {1} (\bibinfo
  {year} {1969})}\BibitemShut {NoStop}%
\bibitem [{\citenamefont {Steiner}\ and\ \citenamefont
  {Gandolfi}(2012)}]{Steiner2012}%
  \BibitemOpen
  \bibfield  {author} {\bibinfo {author} {\bibfnamefont {A.~W.}\ \bibnamefont
  {Steiner}}\ and\ \bibinfo {author} {\bibfnamefont {S.}~\bibnamefont
  {Gandolfi}},\ }\href@noop {} {\bibfield  {journal} {\bibinfo  {journal}
  {Phys. Rev. Lett.}\ }\textbf {\bibinfo {volume} {108}},\ \bibinfo {pages}
  {081102} (\bibinfo {year} {2012})}\BibitemShut {NoStop}%
\bibitem [{\citenamefont {Greiter}\ \emph {et~al.}(1991)\citenamefont
  {Greiter}, \citenamefont {Wen},\ and\ \citenamefont {Wilczek}}]{Greiter1991}%
  \BibitemOpen
  \bibfield  {author} {\bibinfo {author} {\bibfnamefont {M.}~\bibnamefont
  {Greiter}}, \bibinfo {author} {\bibfnamefont {X.-G.}\ \bibnamefont {Wen}}, \
  and\ \bibinfo {author} {\bibfnamefont {F.}~\bibnamefont {Wilczek}},\
  }\href@noop {} {\bibfield  {journal} {\bibinfo  {journal} {Phys. Rev. Lett.}\
  }\textbf {\bibinfo {volume} {66}},\ \bibinfo {pages} {3205} (\bibinfo {year}
  {1991})}\BibitemShut {NoStop}%
\bibitem [{\citenamefont {B{\"u}chler}\ \emph {et~al.}(2007)\citenamefont
  {B{\"u}chler}, \citenamefont {Micheli},\ and\ \citenamefont
  {Zoller}}]{Buchler2007}%
  \BibitemOpen
  \bibfield  {author} {\bibinfo {author} {\bibfnamefont {H.}~\bibnamefont
  {B{\"u}chler}}, \bibinfo {author} {\bibfnamefont {A.}~\bibnamefont
  {Micheli}}, \ and\ \bibinfo {author} {\bibfnamefont {P.}~\bibnamefont
  {Zoller}},\ }\href@noop {} {\bibfield  {journal} {\bibinfo  {journal} {Nat.
  Phys.}\ }\textbf {\bibinfo {volume} {3}},\ \bibinfo {pages} {726} (\bibinfo
  {year} {2007})}\BibitemShut {NoStop}%
\bibitem [{\citenamefont {Johnson}\ \emph {et~al.}(2009)\citenamefont
  {Johnson}, \citenamefont {Tiesinga}, \citenamefont {Porto},\ and\
  \citenamefont {Williams}}]{Johnson2009}%
  \BibitemOpen
  \bibfield  {author} {\bibinfo {author} {\bibfnamefont {P.}~\bibnamefont
  {Johnson}}, \bibinfo {author} {\bibfnamefont {E.}~\bibnamefont {Tiesinga}},
  \bibinfo {author} {\bibfnamefont {J.}~\bibnamefont {Porto}}, \ and\ \bibinfo
  {author} {\bibfnamefont {C.}~\bibnamefont {Williams}},\ }\href@noop {}
  {\bibfield  {journal} {\bibinfo  {journal} {New Journal of Physics}\ }\textbf
  {\bibinfo {volume} {11}},\ \bibinfo {pages} {093022} (\bibinfo {year}
  {2009})}\BibitemShut {NoStop}%
\bibitem [{\citenamefont {Daley}\ \emph {et~al.}(2009)\citenamefont {Daley},
  \citenamefont {Taylor}, \citenamefont {Diehl}, \citenamefont {Baranov},\ and\
  \citenamefont {Zoller}}]{Daley2009}%
  \BibitemOpen
  \bibfield  {author} {\bibinfo {author} {\bibfnamefont {A.~J.}\ \bibnamefont
  {Daley}}, \bibinfo {author} {\bibfnamefont {J.~M.}\ \bibnamefont {Taylor}},
  \bibinfo {author} {\bibfnamefont {S.}~\bibnamefont {Diehl}}, \bibinfo
  {author} {\bibfnamefont {M.}~\bibnamefont {Baranov}}, \ and\ \bibinfo
  {author} {\bibfnamefont {P.}~\bibnamefont {Zoller}},\ }\href@noop {}
  {\bibfield  {journal} {\bibinfo  {journal} {Phys. Rev. Lett.}\ }\textbf
  {\bibinfo {volume} {102}},\ \bibinfo {pages} {040402} (\bibinfo {year}
  {2009})}\BibitemShut {NoStop}%
\bibitem [{\citenamefont {Mazza}\ \emph {et~al.}(2010)\citenamefont {Mazza},
  \citenamefont {Rizzi}, \citenamefont {Lewenstein},\ and\ \citenamefont
  {Cirac}}]{Mazza2010}%
  \BibitemOpen
  \bibfield  {author} {\bibinfo {author} {\bibfnamefont {L.}~\bibnamefont
  {Mazza}}, \bibinfo {author} {\bibfnamefont {M.}~\bibnamefont {Rizzi}},
  \bibinfo {author} {\bibfnamefont {M.}~\bibnamefont {Lewenstein}}, \ and\
  \bibinfo {author} {\bibfnamefont {J.~I.}\ \bibnamefont {Cirac}},\ }\href@noop
  {} {\bibfield  {journal} {\bibinfo  {journal} {Phys. Rev. A}\ }\textbf
  {\bibinfo {volume} {82}},\ \bibinfo {pages} {043629} (\bibinfo {year}
  {2010})}\BibitemShut {NoStop}%
\bibitem [{\citenamefont {Petrov}(2014)}]{Petrov2014}%
  \BibitemOpen
  \bibfield  {author} {\bibinfo {author} {\bibfnamefont {D.}~\bibnamefont
  {Petrov}},\ }\href@noop {} {\bibfield  {journal} {\bibinfo  {journal} {Phys.
  Rev. Lett.}\ }\textbf {\bibinfo {volume} {112}},\ \bibinfo {pages} {103201}
  (\bibinfo {year} {2014})}\BibitemShut {NoStop}%
\bibitem [{\citenamefont {Friedler}\ \emph {et~al.}(2005)\citenamefont
  {Friedler}, \citenamefont {Petrosyan}, \citenamefont {Fleischhauer},\ and\
  \citenamefont {Kurizki}}]{Friedler2005}%
  \BibitemOpen
  \bibfield  {author} {\bibinfo {author} {\bibfnamefont {I.}~\bibnamefont
  {Friedler}}, \bibinfo {author} {\bibfnamefont {D.}~\bibnamefont {Petrosyan}},
  \bibinfo {author} {\bibfnamefont {M.}~\bibnamefont {Fleischhauer}}, \ and\
  \bibinfo {author} {\bibfnamefont {G.}~\bibnamefont {Kurizki}},\ }\href@noop
  {} {\bibfield  {journal} {\bibinfo  {journal} {Phys. Rev. A}\ }\textbf
  {\bibinfo {volume} {72}},\ \bibinfo {pages} {043803} (\bibinfo {year}
  {2005})}\BibitemShut {NoStop}%
\bibitem [{\citenamefont {Gorshkov}\ \emph {et~al.}(2011)\citenamefont
  {Gorshkov}, \citenamefont {Otterbach}, \citenamefont {Fleischhauer},
  \citenamefont {Pohl},\ and\ \citenamefont {Lukin}}]{Gorshkov2011}%
  \BibitemOpen
  \bibfield  {author} {\bibinfo {author} {\bibfnamefont {A.~V.}\ \bibnamefont
  {Gorshkov}}, \bibinfo {author} {\bibfnamefont {J.}~\bibnamefont {Otterbach}},
  \bibinfo {author} {\bibfnamefont {M.}~\bibnamefont {Fleischhauer}}, \bibinfo
  {author} {\bibfnamefont {T.}~\bibnamefont {Pohl}}, \ and\ \bibinfo {author}
  {\bibfnamefont {M.~D.}\ \bibnamefont {Lukin}},\ }\href@noop {} {\bibfield
  {journal} {\bibinfo  {journal} {Phys. Rev. Lett.}\ }\textbf {\bibinfo
  {volume} {107}},\ \bibinfo {pages} {133602} (\bibinfo {year}
  {2011})}\BibitemShut {NoStop}%
\bibitem [{\citenamefont {Peyronel}\ \emph {et~al.}(2012)\citenamefont
  {Peyronel}, \citenamefont {Firstenberg}, \citenamefont {Liang}, \citenamefont
  {Hofferberth}, \citenamefont {Gorshkov}, \citenamefont {Pohl}, \citenamefont
  {Lukin},\ and\ \citenamefont {Vuleti{\'c}}}]{Peyronel2012}%
  \BibitemOpen
  \bibfield  {author} {\bibinfo {author} {\bibfnamefont {T.}~\bibnamefont
  {Peyronel}}, \bibinfo {author} {\bibfnamefont {O.}~\bibnamefont
  {Firstenberg}}, \bibinfo {author} {\bibfnamefont {Q.-Y.}\ \bibnamefont
  {Liang}}, \bibinfo {author} {\bibfnamefont {S.}~\bibnamefont {Hofferberth}},
  \bibinfo {author} {\bibfnamefont {A.~V.}\ \bibnamefont {Gorshkov}}, \bibinfo
  {author} {\bibfnamefont {T.}~\bibnamefont {Pohl}}, \bibinfo {author}
  {\bibfnamefont {M.~D.}\ \bibnamefont {Lukin}}, \ and\ \bibinfo {author}
  {\bibfnamefont {V.}~\bibnamefont {Vuleti{\'c}}},\ }\href@noop {} {\bibfield
  {journal} {\bibinfo  {journal} {Nature}\ }\textbf {\bibinfo {volume} {488}},\
  \bibinfo {pages} {57} (\bibinfo {year} {2012})}\BibitemShut {NoStop}%
\bibitem [{\citenamefont {Parigi}\ \emph {et~al.}(2012)\citenamefont {Parigi},
  \citenamefont {Bimbard}, \citenamefont {Stanojevic}, \citenamefont
  {Hilliard}, \citenamefont {Nogrette}, \citenamefont {Tualle-Brouri},
  \citenamefont {Ourjoumtsev},\ and\ \citenamefont {Grangier}}]{Parigi2012}%
  \BibitemOpen
  \bibfield  {author} {\bibinfo {author} {\bibfnamefont {V.}~\bibnamefont
  {Parigi}}, \bibinfo {author} {\bibfnamefont {E.}~\bibnamefont {Bimbard}},
  \bibinfo {author} {\bibfnamefont {J.}~\bibnamefont {Stanojevic}}, \bibinfo
  {author} {\bibfnamefont {A.~J.}\ \bibnamefont {Hilliard}}, \bibinfo {author}
  {\bibfnamefont {F.}~\bibnamefont {Nogrette}}, \bibinfo {author}
  {\bibfnamefont {R.}~\bibnamefont {Tualle-Brouri}}, \bibinfo {author}
  {\bibfnamefont {A.}~\bibnamefont {Ourjoumtsev}}, \ and\ \bibinfo {author}
  {\bibfnamefont {P.}~\bibnamefont {Grangier}},\ }\href@noop {} {\bibfield
  {journal} {\bibinfo  {journal} {Phys. Rev. Lett.}\ }\textbf {\bibinfo
  {volume} {109}},\ \bibinfo {pages} {233602} (\bibinfo {year}
  {2012})}\BibitemShut {NoStop}%
\bibitem [{\citenamefont {Firstenberg}\ \emph {et~al.}(2013)\citenamefont
  {Firstenberg}, \citenamefont {Peyronel}, \citenamefont {Liang}, \citenamefont
  {Gorshkov}, \citenamefont {Lukin},\ and\ \citenamefont
  {Vuleti{\'c}}}]{Firstenberg2013}%
  \BibitemOpen
  \bibfield  {author} {\bibinfo {author} {\bibfnamefont {O.}~\bibnamefont
  {Firstenberg}}, \bibinfo {author} {\bibfnamefont {T.}~\bibnamefont
  {Peyronel}}, \bibinfo {author} {\bibfnamefont {Q.-Y.}\ \bibnamefont {Liang}},
  \bibinfo {author} {\bibfnamefont {A.~V.}\ \bibnamefont {Gorshkov}}, \bibinfo
  {author} {\bibfnamefont {M.~D.}\ \bibnamefont {Lukin}}, \ and\ \bibinfo
  {author} {\bibfnamefont {V.}~\bibnamefont {Vuleti{\'c}}},\ }\href@noop {}
  {\bibfield  {journal} {\bibinfo  {journal} {Nature}\ }\textbf {\bibinfo
  {volume} {502}},\ \bibinfo {pages} {71} (\bibinfo {year} {2013})}\BibitemShut
  {NoStop}%
\bibitem [{\citenamefont {Fleischhauer}\ \emph {et~al.}(2005)\citenamefont
  {Fleischhauer}, \citenamefont {Imamoglu},\ and\ \citenamefont
  {Marangos}}]{Fleischhauer2005}%
  \BibitemOpen
  \bibfield  {author} {\bibinfo {author} {\bibfnamefont {M.}~\bibnamefont
  {Fleischhauer}}, \bibinfo {author} {\bibfnamefont {A.}~\bibnamefont
  {Imamoglu}}, \ and\ \bibinfo {author} {\bibfnamefont {J.~P.}\ \bibnamefont
  {Marangos}},\ }\href@noop {} {\bibfield  {journal} {\bibinfo  {journal} {Rev.
  Mod. Phys.}\ }\textbf {\bibinfo {volume} {77}},\ \bibinfo {pages} {633}
  (\bibinfo {year} {2005})}\BibitemShut {NoStop}%
\bibitem [{\citenamefont {Fleischhauer}\ and\ \citenamefont
  {Lukin}(2000)}]{Fleischhauer2000}%
  \BibitemOpen
  \bibfield  {author} {\bibinfo {author} {\bibfnamefont {M.}~\bibnamefont
  {Fleischhauer}}\ and\ \bibinfo {author} {\bibfnamefont {M.}~\bibnamefont
  {Lukin}},\ }\href@noop {} {\bibfield  {journal} {\bibinfo  {journal} {Phys.
  Rev. Lett.}\ }\textbf {\bibinfo {volume} {84}},\ \bibinfo {pages} {5094}
  (\bibinfo {year} {2000})}\BibitemShut {NoStop}%
\bibitem [{\citenamefont {Lukin}\ \emph {et~al.}(2001)\citenamefont {Lukin},
  \citenamefont {Fleischhauer}, \citenamefont {Cote}, \citenamefont {Duan},
  \citenamefont {Jaksch}, \citenamefont {Cirac},\ and\ \citenamefont
  {Zoller}}]{Lukin2001}%
  \BibitemOpen
  \bibfield  {author} {\bibinfo {author} {\bibfnamefont {M.~D.}\ \bibnamefont
  {Lukin}}, \bibinfo {author} {\bibfnamefont {M.}~\bibnamefont {Fleischhauer}},
  \bibinfo {author} {\bibfnamefont {R.}~\bibnamefont {Cote}}, \bibinfo {author}
  {\bibfnamefont {L.~M.}\ \bibnamefont {Duan}}, \bibinfo {author}
  {\bibfnamefont {D.}~\bibnamefont {Jaksch}}, \bibinfo {author} {\bibfnamefont
  {J.~I.}\ \bibnamefont {Cirac}}, \ and\ \bibinfo {author} {\bibfnamefont
  {P.}~\bibnamefont {Zoller}},\ }\href@noop {} {\bibfield  {journal} {\bibinfo
  {journal} {Phys. Rev. Lett.}\ }\textbf {\bibinfo {volume} {87}},\ \bibinfo
  {pages} {037901} (\bibinfo {year} {2001})}\BibitemShut {NoStop}%
\bibitem [{\citenamefont {Heidemann}\ \emph {et~al.}(2007)\citenamefont
  {Heidemann}, \citenamefont {Raitzsch}, \citenamefont {Bendkowsky},
  \citenamefont {Butscher}, \citenamefont {L\"ow}, \citenamefont {Santos},\
  and\ \citenamefont {Pfau}}]{Heidemann2007}%
  \BibitemOpen
  \bibfield  {author} {\bibinfo {author} {\bibfnamefont {R.}~\bibnamefont
  {Heidemann}}, \bibinfo {author} {\bibfnamefont {U.}~\bibnamefont {Raitzsch}},
  \bibinfo {author} {\bibfnamefont {V.}~\bibnamefont {Bendkowsky}}, \bibinfo
  {author} {\bibfnamefont {B.}~\bibnamefont {Butscher}}, \bibinfo {author}
  {\bibfnamefont {R.}~\bibnamefont {L\"ow}}, \bibinfo {author} {\bibfnamefont
  {L.}~\bibnamefont {Santos}}, \ and\ \bibinfo {author} {\bibfnamefont
  {T.}~\bibnamefont {Pfau}},\ }\href@noop {} {\bibfield  {journal} {\bibinfo
  {journal} {Phys. Rev. Lett.}\ }\textbf {\bibinfo {volume} {99}},\ \bibinfo
  {pages} {163601} (\bibinfo {year} {2007})}\BibitemShut {NoStop}%
\bibitem [{\citenamefont {Bienias}\ \emph {et~al.}(2014)\citenamefont
  {Bienias}, \citenamefont {Choi}, \citenamefont {Firstenberg}, \citenamefont
  {Maghrebi}, \citenamefont {Gullans}, \citenamefont {Lukin}, \citenamefont
  {Gorshkov},\ and\ \citenamefont {B{\"u}chler}}]{Bienias2014}%
  \BibitemOpen
  \bibfield  {author} {\bibinfo {author} {\bibfnamefont {P.}~\bibnamefont
  {Bienias}}, \bibinfo {author} {\bibfnamefont {S.}~\bibnamefont {Choi}},
  \bibinfo {author} {\bibfnamefont {O.}~\bibnamefont {Firstenberg}}, \bibinfo
  {author} {\bibfnamefont {M.}~\bibnamefont {Maghrebi}}, \bibinfo {author}
  {\bibfnamefont {M.}~\bibnamefont {Gullans}}, \bibinfo {author} {\bibfnamefont
  {M.~D.}\ \bibnamefont {Lukin}}, \bibinfo {author} {\bibfnamefont {A.~V.}\
  \bibnamefont {Gorshkov}}, \ and\ \bibinfo {author} {\bibfnamefont
  {H.}~\bibnamefont {B{\"u}chler}},\ }\href@noop {} {\bibfield  {journal}
  {\bibinfo  {journal} {Phys. Rev. A}\ }\textbf {\bibinfo {volume} {90}},\
  \bibinfo {pages} {053804} (\bibinfo {year} {2014})}\BibitemShut {NoStop}%
\bibitem [{\citenamefont {Maghrebi}\ \emph
  {et~al.}(2015{\natexlab{a}})\citenamefont {Maghrebi}, \citenamefont
  {Gullans}, \citenamefont {Bienias}, \citenamefont {Choi}, \citenamefont
  {Martin}, \citenamefont {Firstenberg}, \citenamefont {Lukin}, \citenamefont
  {B{\"u}chler},\ and\ \citenamefont {Gorshkov}}]{Maghrebi2015}%
  \BibitemOpen
  \bibfield  {author} {\bibinfo {author} {\bibfnamefont {M.}~\bibnamefont
  {Maghrebi}}, \bibinfo {author} {\bibfnamefont {M.}~\bibnamefont {Gullans}},
  \bibinfo {author} {\bibfnamefont {P.}~\bibnamefont {Bienias}}, \bibinfo
  {author} {\bibfnamefont {S.}~\bibnamefont {Choi}}, \bibinfo {author}
  {\bibfnamefont {I.}~\bibnamefont {Martin}}, \bibinfo {author} {\bibfnamefont
  {O.}~\bibnamefont {Firstenberg}}, \bibinfo {author} {\bibfnamefont
  {M.}~\bibnamefont {Lukin}}, \bibinfo {author} {\bibfnamefont
  {H.}~\bibnamefont {B{\"u}chler}}, \ and\ \bibinfo {author} {\bibfnamefont
  {A.}~\bibnamefont {Gorshkov}},\ }\href@noop {} {\bibfield  {journal}
  {\bibinfo  {journal} {Phys. Rev. Lett.}\ }\textbf {\bibinfo {volume} {115}},\
  \bibinfo {pages} {123601} (\bibinfo {year} {2015}{\natexlab{a}})}\BibitemShut
  {NoStop}%
\bibitem [{\citenamefont {Dudin}\ and\ \citenamefont
  {Kuzmich}(2012)}]{Dudin2012}%
  \BibitemOpen
  \bibfield  {author} {\bibinfo {author} {\bibfnamefont {Y.}~\bibnamefont
  {Dudin}}\ and\ \bibinfo {author} {\bibfnamefont {A.}~\bibnamefont
  {Kuzmich}},\ }\href@noop {} {\bibfield  {journal} {\bibinfo  {journal}
  {Science}\ }\textbf {\bibinfo {volume} {336}},\ \bibinfo {pages} {887}
  (\bibinfo {year} {2012})}\BibitemShut {NoStop}%
\bibitem [{\citenamefont {Maxwell}\ \emph {et~al.}(2013)\citenamefont
  {Maxwell}, \citenamefont {Szwer}, \citenamefont {Paredes-Barato},
  \citenamefont {Busche}, \citenamefont {Pritchard}, \citenamefont {Gauguet},
  \citenamefont {Weatherill}, \citenamefont {Jones},\ and\ \citenamefont
  {Adams}}]{Maxwell2013}%
  \BibitemOpen
  \bibfield  {author} {\bibinfo {author} {\bibfnamefont {D.}~\bibnamefont
  {Maxwell}}, \bibinfo {author} {\bibfnamefont {D.~J.}\ \bibnamefont {Szwer}},
  \bibinfo {author} {\bibfnamefont {D.}~\bibnamefont {Paredes-Barato}},
  \bibinfo {author} {\bibfnamefont {H.}~\bibnamefont {Busche}}, \bibinfo
  {author} {\bibfnamefont {J.~D.}\ \bibnamefont {Pritchard}}, \bibinfo {author}
  {\bibfnamefont {A.}~\bibnamefont {Gauguet}}, \bibinfo {author} {\bibfnamefont
  {K.~J.}\ \bibnamefont {Weatherill}}, \bibinfo {author} {\bibfnamefont
  {M.~P.~A.}\ \bibnamefont {Jones}}, \ and\ \bibinfo {author} {\bibfnamefont
  {C.~S.}\ \bibnamefont {Adams}},\ }\href@noop {} {\bibfield  {journal}
  {\bibinfo  {journal} {Phys. Rev. Lett.}\ }\textbf {\bibinfo {volume} {110}},\
  \bibinfo {pages} {103001} (\bibinfo {year} {2013})}\BibitemShut {NoStop}%
\bibitem [{\citenamefont {Hofmann}\ \emph {et~al.}(2013)\citenamefont
  {Hofmann}, \citenamefont {G\"unter}, \citenamefont {Schempp}, \citenamefont
  {Robert-de Saint-Vincent}, \citenamefont {G\"arttner}, \citenamefont {Evers},
  \citenamefont {Whitlock},\ and\ \citenamefont {Weidem\"uller}}]{Hoffman2013}%
  \BibitemOpen
  \bibfield  {author} {\bibinfo {author} {\bibfnamefont {C.~S.}\ \bibnamefont
  {Hofmann}}, \bibinfo {author} {\bibfnamefont {G.}~\bibnamefont {G\"unter}},
  \bibinfo {author} {\bibfnamefont {H.}~\bibnamefont {Schempp}}, \bibinfo
  {author} {\bibfnamefont {M.}~\bibnamefont {Robert-de Saint-Vincent}},
  \bibinfo {author} {\bibfnamefont {M.}~\bibnamefont {G\"arttner}}, \bibinfo
  {author} {\bibfnamefont {J.}~\bibnamefont {Evers}}, \bibinfo {author}
  {\bibfnamefont {S.}~\bibnamefont {Whitlock}}, \ and\ \bibinfo {author}
  {\bibfnamefont {M.}~\bibnamefont {Weidem\"uller}},\ }\href@noop {} {\bibfield
   {journal} {\bibinfo  {journal} {Phys. Rev. Lett.}\ }\textbf {\bibinfo
  {volume} {110}},\ \bibinfo {pages} {203601} (\bibinfo {year}
  {2013})}\BibitemShut {NoStop}%
\bibitem [{\citenamefont {Gorniaczyk}\ \emph {et~al.}(2014)\citenamefont
  {Gorniaczyk}, \citenamefont {Tresp}, \citenamefont {Schmidt}, \citenamefont
  {Fedder},\ and\ \citenamefont {Hofferberth}}]{Gorniaczyk2014}%
  \BibitemOpen
  \bibfield  {author} {\bibinfo {author} {\bibfnamefont {H.}~\bibnamefont
  {Gorniaczyk}}, \bibinfo {author} {\bibfnamefont {C.}~\bibnamefont {Tresp}},
  \bibinfo {author} {\bibfnamefont {J.}~\bibnamefont {Schmidt}}, \bibinfo
  {author} {\bibfnamefont {H.}~\bibnamefont {Fedder}}, \ and\ \bibinfo {author}
  {\bibfnamefont {S.}~\bibnamefont {Hofferberth}},\ }\href@noop {} {\bibfield
  {journal} {\bibinfo  {journal} {Phys. Rev. Lett.}\ }\textbf {\bibinfo
  {volume} {113}},\ \bibinfo {pages} {053601} (\bibinfo {year}
  {2014})}\BibitemShut {NoStop}%
\bibitem [{\citenamefont {Tiarks}\ \emph {et~al.}(2014)\citenamefont {Tiarks},
  \citenamefont {Baur}, \citenamefont {Schneider}, \citenamefont {D\"urr},\
  and\ \citenamefont {Rempe}}]{Tiarks2014}%
  \BibitemOpen
  \bibfield  {author} {\bibinfo {author} {\bibfnamefont {D.}~\bibnamefont
  {Tiarks}}, \bibinfo {author} {\bibfnamefont {S.}~\bibnamefont {Baur}},
  \bibinfo {author} {\bibfnamefont {K.}~\bibnamefont {Schneider}}, \bibinfo
  {author} {\bibfnamefont {S.}~\bibnamefont {D\"urr}}, \ and\ \bibinfo {author}
  {\bibfnamefont {G.}~\bibnamefont {Rempe}},\ }\href@noop {} {\bibfield
  {journal} {\bibinfo  {journal} {Phys. Rev. Lett.}\ }\textbf {\bibinfo
  {volume} {113}},\ \bibinfo {pages} {053602} (\bibinfo {year}
  {2014})}\BibitemShut {NoStop}%
\bibitem [{\citenamefont {Otterbach}\ \emph {et~al.}(2013)\citenamefont
  {Otterbach}, \citenamefont {Moos}, \citenamefont {Muth},\ and\ \citenamefont
  {Fleischhauer}}]{Otterbach2013}%
  \BibitemOpen
  \bibfield  {author} {\bibinfo {author} {\bibfnamefont {J.}~\bibnamefont
  {Otterbach}}, \bibinfo {author} {\bibfnamefont {M.}~\bibnamefont {Moos}},
  \bibinfo {author} {\bibfnamefont {D.}~\bibnamefont {Muth}}, \ and\ \bibinfo
  {author} {\bibfnamefont {M.}~\bibnamefont {Fleischhauer}},\ }\href@noop {}
  {\bibfield  {journal} {\bibinfo  {journal} {Phys. Rev. Lett.}\ }\textbf
  {\bibinfo {volume} {111}},\ \bibinfo {pages} {113001} (\bibinfo {year}
  {2013})}\BibitemShut {NoStop}%
\bibitem [{\citenamefont {Gorshkov}\ \emph {et~al.}(2013)\citenamefont
  {Gorshkov}, \citenamefont {Nath},\ and\ \citenamefont {Pohl}}]{Gorshkov2013}%
  \BibitemOpen
  \bibfield  {author} {\bibinfo {author} {\bibfnamefont {A.~V.}\ \bibnamefont
  {Gorshkov}}, \bibinfo {author} {\bibfnamefont {R.}~\bibnamefont {Nath}}, \
  and\ \bibinfo {author} {\bibfnamefont {T.}~\bibnamefont {Pohl}},\ }\href@noop
  {} {\bibfield  {journal} {\bibinfo  {journal} {Phys. Rev. Lett.}\ }\textbf
  {\bibinfo {volume} {110}},\ \bibinfo {pages} {153601} (\bibinfo {year}
  {2013})}\BibitemShut {NoStop}%
\bibitem [{\citenamefont {Maghrebi}\ \emph
  {et~al.}(2015{\natexlab{b}})\citenamefont {Maghrebi}, \citenamefont {Yao},
  \citenamefont {Hafezi}, \citenamefont {Pohl}, \citenamefont {Firstenberg},\
  and\ \citenamefont {Gorshkov}}]{Maghrebi2015a}%
  \BibitemOpen
  \bibfield  {author} {\bibinfo {author} {\bibfnamefont {M.~F.}\ \bibnamefont
  {Maghrebi}}, \bibinfo {author} {\bibfnamefont {N.~Y.}\ \bibnamefont {Yao}},
  \bibinfo {author} {\bibfnamefont {M.}~\bibnamefont {Hafezi}}, \bibinfo
  {author} {\bibfnamefont {T.}~\bibnamefont {Pohl}}, \bibinfo {author}
  {\bibfnamefont {O.}~\bibnamefont {Firstenberg}}, \ and\ \bibinfo {author}
  {\bibfnamefont {A.~V.}\ \bibnamefont {Gorshkov}},\ }\href@noop {} {\bibfield
  {journal} {\bibinfo  {journal} {Phys. Rev. A}\ }\textbf {\bibinfo {volume}
  {91}},\ \bibinfo {pages} {033838} (\bibinfo {year}
  {2015}{\natexlab{b}})}\BibitemShut {NoStop}%
\bibitem [{\citenamefont {Moos}\ \emph {et~al.}(2015)\citenamefont {Moos},
  \citenamefont {H{\"o}ning}, \citenamefont {Unanyan},\ and\ \citenamefont
  {Fleischhauer}}]{Moos2015}%
  \BibitemOpen
  \bibfield  {author} {\bibinfo {author} {\bibfnamefont {M.}~\bibnamefont
  {Moos}}, \bibinfo {author} {\bibfnamefont {M.}~\bibnamefont {H{\"o}ning}},
  \bibinfo {author} {\bibfnamefont {R.}~\bibnamefont {Unanyan}}, \ and\
  \bibinfo {author} {\bibfnamefont {M.}~\bibnamefont {Fleischhauer}},\
  }\href@noop {} {\bibfield  {journal} {\bibinfo  {journal} {Phys. Rev. A}\
  }\textbf {\bibinfo {volume} {92}},\ \bibinfo {pages} {053846} (\bibinfo
  {year} {2015})}\BibitemShut {NoStop}%
\bibitem [{\citenamefont {Weber}\ \emph {et~al.}(2015)\citenamefont {Weber},
  \citenamefont {H{\"o}ning}, \citenamefont {Niederpr{\"u}m}, \citenamefont
  {Manthey}, \citenamefont {Thomas}, \citenamefont {Guarrera}, \citenamefont
  {Fleischhauer}, \citenamefont {Barontini},\ and\ \citenamefont
  {Ott}}]{Weber2015}%
  \BibitemOpen
  \bibfield  {author} {\bibinfo {author} {\bibfnamefont {T.}~\bibnamefont
  {Weber}}, \bibinfo {author} {\bibfnamefont {M.}~\bibnamefont {H{\"o}ning}},
  \bibinfo {author} {\bibfnamefont {T.}~\bibnamefont {Niederpr{\"u}m}},
  \bibinfo {author} {\bibfnamefont {T.}~\bibnamefont {Manthey}}, \bibinfo
  {author} {\bibfnamefont {O.}~\bibnamefont {Thomas}}, \bibinfo {author}
  {\bibfnamefont {V.}~\bibnamefont {Guarrera}}, \bibinfo {author}
  {\bibfnamefont {M.}~\bibnamefont {Fleischhauer}}, \bibinfo {author}
  {\bibfnamefont {G.}~\bibnamefont {Barontini}}, \ and\ \bibinfo {author}
  {\bibfnamefont {H.}~\bibnamefont {Ott}},\ }\href@noop {} {\bibfield
  {journal} {\bibinfo  {journal} {Nat. Phys.}\ } (\bibinfo {year}
  {2015})}\BibitemShut {NoStop}%
\bibitem [{\citenamefont {Sommer}\ \emph {et~al.}(2015)\citenamefont {Sommer},
  \citenamefont {B{\"u}chler},\ and\ \citenamefont {Simon}}]{Sommer2015}%
  \BibitemOpen
  \bibfield  {author} {\bibinfo {author} {\bibfnamefont {A.}~\bibnamefont
  {Sommer}}, \bibinfo {author} {\bibfnamefont {H.~P.}\ \bibnamefont
  {B{\"u}chler}}, \ and\ \bibinfo {author} {\bibfnamefont {J.}~\bibnamefont
  {Simon}},\ }\href@noop {} {\bibfield  {journal} {\bibinfo  {journal} {arXiv
  preprint arXiv:1506.00341}\ } (\bibinfo {year} {2015})}\BibitemShut {NoStop}%
\bibitem [{\citenamefont {Ningyuan}\ \emph {et~al.}(2015)\citenamefont
  {Ningyuan}, \citenamefont {Georgakopoulos}, \citenamefont {Ryou},
  \citenamefont {Schine}, \citenamefont {Sommer},\ and\ \citenamefont
  {Simon}}]{Ningyuan2015}%
  \BibitemOpen
  \bibfield  {author} {\bibinfo {author} {\bibfnamefont {J.}~\bibnamefont
  {Ningyuan}}, \bibinfo {author} {\bibfnamefont {A.}~\bibnamefont
  {Georgakopoulos}}, \bibinfo {author} {\bibfnamefont {A.}~\bibnamefont
  {Ryou}}, \bibinfo {author} {\bibfnamefont {N.}~\bibnamefont {Schine}},
  \bibinfo {author} {\bibfnamefont {A.}~\bibnamefont {Sommer}}, \ and\ \bibinfo
  {author} {\bibfnamefont {J.}~\bibnamefont {Simon}},\ }\href@noop {}
  {\bibfield  {journal} {\bibinfo  {journal} {arXiv preprint arXiv:1511.01872}\
  } (\bibinfo {year} {2015})}\BibitemShut {NoStop}%
\bibitem [{Sup()}]{SupMat}%
  \BibitemOpen
  \href@noop {} {}\bibinfo {note} {See Supplemental Material, which contains
  references~\cite{Gibson1987,Dincao2014,Trefethen2000}, for detailed
  derivation of the three-body interaction, description of the adiabatic
  approach and numerical details.}\BibitemShut {Stop}%
\bibitem [{\citenamefont {McGuire}(1964)}]{Mcguire1964}%
  \BibitemOpen
  \bibfield  {author} {\bibinfo {author} {\bibfnamefont {J.~B.}\ \bibnamefont
  {McGuire}},\ }\href@noop {} {\bibfield  {journal} {\bibinfo  {journal}
  {Journal of Mathematical Physics}\ }\textbf {\bibinfo {volume} {5}},\
  \bibinfo {pages} {622} (\bibinfo {year} {1964})}\BibitemShut {NoStop}%
\bibitem [{\citenamefont {Gibson}\ \emph {et~al.}(1987)\citenamefont {Gibson},
  \citenamefont {Larsen},\ and\ \citenamefont {Popiel}}]{Gibson1987}%
  \BibitemOpen
  \bibfield  {author} {\bibinfo {author} {\bibfnamefont {W.~G.}\ \bibnamefont
  {Gibson}}, \bibinfo {author} {\bibfnamefont {S.~Y.}\ \bibnamefont {Larsen}},
  \ and\ \bibinfo {author} {\bibfnamefont {J.}~\bibnamefont {Popiel}},\
  }\href@noop {} {\bibfield  {journal} {\bibinfo  {journal} {Phys. Rev. A}\
  }\textbf {\bibinfo {volume} {35}},\ \bibinfo {pages} {4919} (\bibinfo {year}
  {1987})}\BibitemShut {NoStop}%
\bibitem [{\citenamefont {Mehta}\ and\ \citenamefont
  {Shepard}(2005)}]{Mehta2005}%
  \BibitemOpen
  \bibfield  {author} {\bibinfo {author} {\bibfnamefont {N.~P.}\ \bibnamefont
  {Mehta}}\ and\ \bibinfo {author} {\bibfnamefont {J.~R.}\ \bibnamefont
  {Shepard}},\ }\href@noop {} {\bibfield  {journal} {\bibinfo  {journal} {Phys.
  Rev. A}\ }\textbf {\bibinfo {volume} {72}},\ \bibinfo {pages} {032728}
  (\bibinfo {year} {2005})}\BibitemShut {NoStop}%
\bibitem [{\citenamefont {Gullans}\ \emph {et~al.}(2016)\citenamefont
  {Gullans}, \citenamefont {Wang}, \citenamefont {Thompson}, \citenamefont
  {Liang}, \citenamefont {Vuletic}, \citenamefont {Lukin},\ and\ \citenamefont
  {Gorshkov}}]{Gullans2016}%
  \BibitemOpen
  \bibfield  {author} {\bibinfo {author} {\bibfnamefont {M.}~\bibnamefont
  {Gullans}}, \bibinfo {author} {\bibfnamefont {Y.}~\bibnamefont {Wang}},
  \bibinfo {author} {\bibfnamefont {J.}~\bibnamefont {Thompson}}, \bibinfo
  {author} {\bibfnamefont {Q.-Y.}\ \bibnamefont {Liang}}, \bibinfo {author}
  {\bibfnamefont {V.}~\bibnamefont {Vuletic}}, \bibinfo {author} {\bibfnamefont
  {M.}~\bibnamefont {Lukin}}, \ and\ \bibinfo {author} {\bibfnamefont
  {A.}~\bibnamefont {Gorshkov}},\ }\href@noop {} {\bibfield  {journal}
  {\bibinfo  {journal} {arXiv preprint arXiv:1605.05651}\ } (\bibinfo {year}
  {2016})}\BibitemShut {NoStop}%
\bibitem [{\citenamefont {D'Incao}\ and\ \citenamefont
  {Esry}(2014)}]{Dincao2014}%
  \BibitemOpen
  \bibfield  {author} {\bibinfo {author} {\bibfnamefont {J.~P.}\ \bibnamefont
  {D'Incao}}\ and\ \bibinfo {author} {\bibfnamefont {B.~D.}\ \bibnamefont
  {Esry}},\ }\href@noop {} {\bibfield  {journal} {\bibinfo  {journal} {Phys.
  Rev. A}\ }\textbf {\bibinfo {volume} {90}},\ \bibinfo {pages} {042707}
  (\bibinfo {year} {2014})}\BibitemShut {NoStop}%
\bibitem [{\citenamefont {Trefethen}(2000)}]{Trefethen2000}%
  \BibitemOpen
  \bibfield  {author} {\bibinfo {author} {\bibfnamefont {L.~N.}\ \bibnamefont
  {Trefethen}},\ }\href@noop {} {\emph {\bibinfo {title} {Spectral methods in
  MATLAB}}},\ Vol.~\bibinfo {volume} {10}\ (\bibinfo  {publisher} {Siam},\
  \bibinfo {year} {2000})\BibitemShut {NoStop}%
\end{thebibliography}%

\clearpage


\section*{Supplementary material}

# Supplementary Material to "Three-body interactions of slow light Rydberg polaritons"

Krzysztof Jachymski[1], Przemyslaw Bienias[1] and Hans Peter Büchler[1]

*Institute for Theoretical Physics III and Center for Integrated Quantum Science and Technology,*
*University of Stuttgart, Pfaffenwaldring 57, 70550 Stuttgart, Germany*

(Dated: June 15, 2016)

## I. TWO-BODY INTERACTION

In the first part, we derive Eq. (1) for the two body interaction from the main text. We denote the function describing the photonic mode by $h(x)$. The stationary Schrödinger equation for the different parts with two, one and zero photons can then be written as

$$\omega\phi_0 = -2\nu\phi_0 + \int dx\, \phi_1(x)h^*(x), \tag{1}$$

$$\omega\phi_1(x) = -(\nu + \frac{1}{\nu})\phi_1(x) + 2h(x)\phi_0 + 2\int dy\, \phi_2(x,y)h^*(y), \tag{2}$$

$$\omega\phi_2(x,y) = -\frac{2}{\nu}\phi_2(x,y) + \frac{1}{2}(h(x)\phi_1(y) + h(y)\phi_1(x)) + V(x-y)\phi_2(x,y). \tag{3}$$

Here the equations are given in units of $g\Omega/\Delta$, which naturally appear in the coupling terms. The photonic mode is normalized $\int dx|h(x)|^2 = 1$, and we introduce the notation $\nu = g/\Omega$. Without loss of generality, we set $\phi_0 = 1$. The last equation provides

$$\phi_2(x,y) = \frac{h(x)\phi_1(y) + h(y)\phi_1(x)}{2(\omega + 2/\nu)}\left[1 + \frac{(\omega + 2/\nu)^{-1}V(x-y)}{1 - (\omega + 2/\nu)^{-1}V(x-y)}\right]. \tag{4}$$

Inserting this result into the second equation, multiplying by $h^*(x)$, and performing the integration, we obtain

$$(\omega + \frac{2}{\nu})(\omega + \nu + \frac{1}{\nu})(\omega + 2\nu) - 2(\omega + \frac{2}{\nu}) - 2(\omega + 2\nu) = 2\int dxdy \frac{(\omega + 2/\nu)^{-1}V(x-y)}{1 - (\omega + 2/\nu)^{-1}V(x-y)}|h(x)|^2 h^*(y)\phi_1(y). \tag{5}$$

Here, we have used the first equation to replace $\int dx\phi_1(x)h(x)$ by $\omega + 2\nu$. The right side is a small contribution for a cavity mode much larger than the blockade radius. Therefore, expanding the left hand side for small $\omega$, we obtain the energy shift

$$\omega = \frac{\nu^3}{2(1 + \nu^2)^2}\int dxdy \frac{V(x-y)}{1 - \nu/2 V(x-y)}|h(x)|^2 h^*(x)\phi_1(x). \tag{6}$$

In leading order, $\phi_1(x)$ reduces to $\phi_1 \approx h(x)2\nu$ and we recover the effective interaction potential from the main text.

## II. THREE-BODY INTERACTION

Next, we can derive Eq. (2) from the main text for the three-body interaction. Again, we denote the function describing the photonic mode by $h(x)$, and the total wave function can be decomposed into parts containing three, two, one and zero atomic excitations, denoted as $\phi_i$. The stationary Schrödinger equation can then be written as

$$\omega\phi_0 = -3\nu\phi_0 + \int dx\, \phi_1(x)h^*(x), \tag{7}$$

$$\omega\phi_1(x) = -(2\nu + \frac{1}{\nu})\phi_1(x) + 3h(x)\phi_0 + 2\int dy\, \phi_2(x,y)h^*(y), \tag{8}$$

$$\omega\phi_2(x,y) = -(\nu + \frac{2}{\nu})\phi_2(x,y) + \frac{1}{2}(h(x)\phi_1(y) + h(y)\phi_1(x)) + 3\int dz\, \phi_3(x,y,z)h^*(z) + V(x-y)\phi_2(x,y), \tag{9}$$

$$\omega\phi_3(x,y,z) = -\frac{3}{\nu}\phi_3(x,y,z) + \frac{1}{3}(h(z)\phi_2(x,y) + h(y)\phi_2(x,z) + h(x)\phi_2(y,z)) + W(x,y,z)\phi_3(x,y,z). \tag{10}$$

Here we again use units of $g\Omega/\Delta$, normalize the photonic mode as $\int dx |h(x)|^2 = 1$, and introduce the notation $\nu = g/\Omega$. The interaction between three Rydberg atoms is determined by the sum of two-body interactions $W(x, y, z) = V(x - y) + V(y - z) + V(z - x)$. We are here interested in the case where the photonic mode is much larger in size than the blockade radius. We fix the normalization of the total wave function by setting $\phi_0 = 1$. Eq. (7) allows us to replace terms $\int dx \phi_1(x) h(x)$ by $\omega + 3\nu$. As a next step, we express $\phi_2$ in terms of an arbitrary function $u(x, y)$ via

$$\phi_2(x, y) = \beta h(x) h(y) \left(1 + \frac{u(x, y)}{1 - u(x, y)}\right). \tag{11}$$

This notation is motivated by the solution for two-polaritons and does not provide any restriction on the general solution. We expect that $\frac{u(x,y)}{1-u(x,y)}$ accounts for the local modification to the shape due to the strong interaction. In the absence of interactions, one would have $u(x, y) = 0$ and $\beta = 3\nu^2$. We thus write $\beta = 3\nu^2(1 + t)$ with $t$ expected to be a small correction. Now we multiply Eq. (8) by $h^*(x)$ and integrate over $x$, which provides the simplified equation

$$\left(\omega + 2\nu + \frac{1}{\nu}\right)(\omega + 3\nu) = 3 + 6\nu^2 \int dx\, dy\, |h(x)|^2 |h(y)|^2 (1 + t) \left(1 + \frac{u(x, y)}{1 - u(x, y)}\right). \tag{12}$$

Here $I = \int dx\, dy\, h(x)^2 h(y)^2 \left(\frac{u(x,y)}{1-u(x,y)}\right)$ is also expected to be a small correction. Up to first order in $t$, $I$ and $\omega$ we therefore obtain the relation

$$\omega\left(5\nu + \frac{1}{\nu}\right) = 6\nu^2(t + I). \tag{13}$$

In the next step, we solve Eq. (10) for $\phi_3$ and insert it into (9), multiply it by $h^*(x)h^*(y)$ and integrate over the variables. The resulting expression is

$$\left(\omega + \nu + \frac{2}{\nu}\right) 3\nu^2(1 + t)(1 + I) = 2(\omega + 3\nu) + 3\nu^2(1 + t) \int dx dy |h(x)|^2 |h(y)|^2 V(x - y)\left(1 + \frac{u(x, y)}{1 - u(x, y)}\right) +$$
$$+ \frac{9\nu^2(1 + t)}{\omega + \frac{3}{\nu}} \left(1 + I + \frac{1}{\omega + \frac{3}{\nu}} \int dx dy dz |h(x)|^2 |h(y)|^2 |h(z)|^2 \frac{V_3(x, y, z)}{1 - u(x, y)}\right), \tag{14}$$

where

$$V_3(x, y, z) = \frac{W(x, y, z)}{1 - \frac{1}{\omega + (3/\nu)} W(x, y, z)} \tag{15}$$

denotes the renormalized interaction term that naturally derives from solving Eq. (10) for $\phi_3$. Expanding Eq. (14) to the first order in small parameters (inserting $t$ from Eq. (13)), we arrive at a surprisingly simple result

$$\omega = \frac{\nu}{(\nu + \frac{1}{\nu})^3} \left(3 \int dx dy dz |h(x)|^2 |h(y)|^2 |h(z)|^2 \frac{V(x - y)}{1 - u(x, y)} + \nu^2 \int dx dy dz |h(x)|^2 |h(y)|^2 |h(z)|^2 \frac{V_3(x, y, z)}{1 - u(x, y)}\right). \tag{16}$$

In the last step, we can finally, determine the form of $u(x, y)$. As the energy shift $\omega$ is already determined in leading order in the small parameters, it is sufficient to determine the shape $u(x, y)$ also in leading order. The latter derives immediately from Eq. (9) and provides the result $u(x, y) = \nu V(x - y)/2$. Therefore, we find that the part with two atomic excitations follows the standard two-particle blockade physics, while the contribution from three excitations is only important for the correction $t$.

Equation (16) contains the contribution to the energy shift $\omega$ from the interactions. Coming back to standard units, we read out the interaction term under the integrals, which takes the form

$$U(x, y, z) = \frac{g^6}{3(g^2 + \Omega^2)^3} \frac{W(x, y, z)}{1 - \frac{\Delta}{3\Omega^2} W(x, y, z)} K(x, y, z) - \frac{g^4 \Omega^2}{(g^2 + \Omega^2)^3} \frac{2\Omega^2}{\Delta} [3 - K(x, y, z)]. \tag{17}$$

with $K(x_1, x_2, x_3) = \sum_{i<j} \left[1 - \frac{\Delta}{2\Omega^2} V(x_i - x_j)\right]^{-1}$. From this we can extract the pure three body term by subtracting the contribution from the two body interactions. The latter takes the form

$$V_{\text{eff}}^{(2)}(x - y) + V_{\text{eff}}^{(2)}(y - z) + V_{\text{eff}}^{(2)}(z - x) = -\frac{g^4}{(g^2 + \Omega^2)^3} \frac{2\Omega^2}{\Delta} [3 - K(x, y, z)]. \tag{18}$$

and therefore, the three-body interaction reduces to

$$\frac{V_{\text{eff}}^{(3)}(x,y,z)}{\alpha^3} = \frac{1}{3} \frac{W(x,y,z)}{1 - \frac{\Delta}{3\Omega^2}W(x,y,z)}K(x,y,z) + \frac{2\Omega^2}{\Delta}\left[3 - K(x,y,z)\right] \tag{19}$$

with $\alpha = g^2/(g^2 + \Omega^2)$. This result can easily be extended to the case of three distinct photonic modes instead of a single common mode $h(x)$. Furthermore, it is clear that the analysis is independent on the dimension of the system.

We note that as expected, the three-body interaction term is proportional to $\alpha^3$, while the two-body scales as $\alpha^2$. Also, if one of the particles is separated from the other two so that two of the terms under each sum in (19) can be neglected, the three-body term reduces strictly to zero.

## III. ADIABATIC POTENTIALS

We will now present in more detail the adiabatic potentials method for the sake of completeness, referring the reader e.g. to [1, 2] for more details. The Jacobi coordinates which appear in the part of the Hamiltonian describing the relative motion are constructed as a relative distance between a chose pair of particles, and an orthogonal vector which can be associated with the distance of the third particle from the centre of mass of the chosen pair. In the first step one transfers from Jacobi to hyperspherical coordinates, which in one dimension are just polar coordinates defined as

$$\eta = \rho\cos\theta, \quad \nu = \rho\sin\theta. \tag{20}$$

Intuitively, $\rho$ accounts for the total size of the three-body system and $\theta$ for its internal structure.

The relative Hamiltonian in hyperspsherical coordinates reads

$$H = -\left(\frac{1}{\rho}\frac{\partial}{\partial\rho}\rho\frac{\partial}{\partial\rho} + \frac{1}{\rho^2}\frac{\partial^2}{\partial\theta^2}\right) + U(\rho,\theta), \tag{21}$$

where $U(\rho,\theta)$ contains the interactions and has a six-fold symmetry. The partial wave expansion $\psi = \sum_k \frac{\Phi_k(\rho)}{\sqrt{\rho}}\frac{\exp(ik\theta)}{\sqrt{2\pi}}$ where k takes values from $-\infty$ to $+\infty$ leads then to a set of equations

$$\left(-\frac{d^2}{d\rho^2} + \frac{k^2 - 1/4}{\rho^2}\right)\Phi_k(\rho) + \sum_{k'} U_{kk'}(\rho)\Phi_{k'}(\rho) = E\Phi_k(\rho), \tag{22}$$

where $U_{kk'}(\rho) = \frac{1}{2\pi}\int e^{i(k'-k)\theta}U(\rho,\theta)d\theta$. Due to the symmetry, only terms with $k - k' = 0 \bmod 6$ do not vanish. We note that at small $\rho$ the interactions that we consider here become isotropic and all the terms with $k \neq k'$ are then small. We can rewrite eq. (22) in the matrix form as

$$\frac{d^2\boldsymbol{\Phi}(\rho)}{d\rho^2} + E\boldsymbol{\Phi}(\rho) - \mathbf{M}(\rho)\boldsymbol{\Phi}(\rho) = 0, \tag{23}$$

where $M_{kk'} = (k^2 - 1/4)\delta_{kk'}/\rho^2 + U_{kk'}$. At each $\rho$ this equation can be diagonalized by a unitary operator $\mathbf{Z}(\rho)$, resulting in diagonal matrix of adiabatic potentials $\boldsymbol{\Lambda}(\rho) = \mathbf{Z}^\dagger(\rho)\mathbf{M}(\rho)\mathbf{Z}(\rho)$. Nonadiabatic terms are proportional to $\frac{dZ}{d\rho}$ and $\frac{d^2Z}{d\rho^2}$ and can be omitted if the diagonalization matrix is slowly varying. In such a case, the equations decouple and each channel can be characterized by effective adiabatic curve $\Lambda_k$, or an effective interaction potential $\Delta_k = \Lambda_k - (k^2 - 1/4)/\rho^2$. Only the lowest $k = 0$ channel is attractive and can support bound states. The threshold energy (the asymptotic energy for $\rho \to \infty$) of this channel corresponds to the opening of atom-dimer continuum, so the bound states of this potential correspond to trimers.

## IV. NUMERICAL CALCULATIONS

The numerical calculations have been performed using Chebyshev spectral method [3]. In this method one maps each spatial coordinate $x_k$ onto $y_k \in [-1, 1]$ range using tangens transformation and solves the Schrödinger equation on a discrete grid constructed using Chebyshev points defined as $y_k^j = \cos(j\pi/N)$, where $N$ is the number of grid points. This provides a convenient global method for finding the lowest eigenstates with high precision.

Estimation of the correlation function $g^{(3)}$ of the outgoing photons is based on a number of approximations analyzed in detail in [4]. In our simplified treatment we neglect the shape of the atomic cloud and simplify the real boundary conditions for three photons entering the medium by a single equation at $R = 0$

$$g^{(3)}(R = 0, \eta, \zeta) = 1.$$

Due to these simplifications we can assume that the evolution of the relative part of the wave function can be described as simple propagation in effective time controlled by $R$. It is then enough to decompose the initial state in the basis of the eigenstates. During the propagation the states pick up phases depending on the energy. It is important to notice that the bound states determine the correlation function at short relative distances, while the scattering states form a diffusive background [5]. We expect that our numerically obtained eigenstates will provide the correct behavior of the correlation function at small $\eta$, $\zeta$ only, because the scattering states cannot be calculated with sufficient precision. We checked that the numerical approach correctly reproduces the peak of the two-body correlation function. As the three-body peak also resembles in shape the structure of the bound states, the bound-state-dominated regions should be correctly accounted for.

---

[1] W. G. Gibson, S. Y. Larsen, and J. Popiel, Phys. Rev. A **35**, 4919 (1987).

[2] J. P. D'Incao and B. D. Esry, Phys. Rev. A **90**, 042707 (2014).

[3] L. N. Trefethen, *Spectral methods in MATLAB*, Vol. 10 (Siam, 2000).

[4] O. Firstenberg, T. Peyronel, Q.-Y. Liang, A. V. Gorshkov, M. D. Lukin, and V. Vuletić, Nature **502**, 71 (2013).

[5] T. Peyronel, O. Firstenberg, Q.-Y. Liang, S. Hofferberth, A. V. Gorshkov, T. Pohl, M. D. Lukin, and V. Vuletić, Nature **488**, 57 (2012).



\end{document}